\documentclass[prl,aps,groupedaddress,floatfix,twocolumn,superscriptaddress,showpacs]{revtex4-1}
\usepackage{amsmath}  
\usepackage{amsfonts} 
\usepackage{graphicx} 
\usepackage{bm}        
\usepackage{amssymb}   
\usepackage{float}
\usepackage{physics}
\usepackage{slashed}
\usepackage{booktabs}
\usepackage{xcolor,colortbl,tabularx}
\usepackage{epstopdf}
\usepackage{bbold}

\newcommand{\beq}{\begin{equation}}
\newcommand{\eeq}{\end{equation}}
\newcommand{\bea}{\begin{eqnarray}}
\newcommand{\eea}{\end{eqnarray}}

\begin{document}

\title{Virial coefficients of 1D and 2D Fermi gases by stochastic methods \\ and a semiclassical lattice approximation}

\author{C. R. Shill}
\author{J. E. Drut}
\affiliation{Department of Physics and Astronomy, University of North Carolina, Chapel Hill, North Carolina 27599, USA}

\preprint{INT-PUB-18-041}

\begin{abstract}
We map out the interaction effects on the first six virial coefficients of one-dimensional Fermi gases with
zero-range attractive and repulsive interactions, and the first four virial coefficients of the two-dimensional
analogue with attractive interactions. To that end, we use two non-perturbative stochastic methods:
projection by complex stochastic quantization, which allows us to determine high-order coefficients at weak coupling
and estimate the radius of convergence of the virial expansion;
and a path-integral representation of the virial coefficients.
To complement our numerical calculations, we present leading-order results in a semiclassical lattice
approximation, which we find to be surprisingly close to the expected answers.
\end{abstract}

\date{\today}

\maketitle 

{\it Introduction}.- The thermodynamics of strongly coupled matter is a topic of current interest in areas of physics that
cover a wide range of scales, from quantum chromodynamics (QCD)~\cite{FiniteDQCD0} to ultracold atoms~\cite{Review1,Review2,ExpReviewLattices}. 
The finite-temperature and density behavior of QCD is, in fact, one of the pressing challenges of that field,
as QCD at finite baryon chemical potential is realized in relativistic heavy-ion collisions and deep inside
neutron stars~\cite{FiniteDQCD0, AstroPhysReview}. On the other hand, ultracold atoms have become an
especially appealing laboratory to probe the properties of strongly coupled matter, due to their purity and malleability, 
and in particular due to the experimentalists' power to modify the interaction
by dialing an external magnetic field across a Feshbach resonance~\cite{ResonancesReview}. Naturally, this amount of control on the
experimental side poses a challenge to theoretical approaches. Indeed, strongly coupled atoms can be routinely 
studied, but their precise quantitative analysis on the theory side usually requires ab initio non-perturbative
tools such as quantum Monte Carlo methods. 

An alternative way to characterize the thermodynamics of a many-body system
has historically been given by the virial expansion (VE), which is non-perturbative and valid in the dilute limit.
The VE is an expansion in powers of the fugacity $z=e^{\beta\mu}$ (where $\beta$ is the inverse temperature 
and $\mu$ is the chemical potential), such that the grand-canonical partition function is written as
\beq
\label{Eq:Zexpansion}
\mathcal Z = \sum_{n=0}^{\infty}  Q_n z^n,
\eeq
where $Q_n$ are the $n$-particle canonical partition functions.
We arrive at the most common form of the VE by expanding the pressure $P$
in powers of $z$:
\beq
\label{Eq:logZexpansion}
\beta P V = \ln \mathcal Z =  Q_1 \sum_{n=1}^{\infty} b_n z^n,
\eeq
where $V$ is the ($d$-dimensional, spatial) volume and $b_n$ are the virial coefficients.
Other quantities of interest besides $P$ can also be expanded in powers of $z$ (see e.g.~\cite{VirialReview}).
The appeal of the VE is that it encodes, at order $n$, 
how the $2$- through $n$-body problems govern the physics of the many-body system. 
Using Eq.~(\ref{Eq:Zexpansion}) in Eq.~(\ref{Eq:logZexpansion}) one sees this explicitly:
\bea
b_2 &=& \frac{Q_2}{Q_1} - \frac{Q_1}{2}, \\
b_3 &=& \frac{Q_3}{Q_1} - Q_2 + \frac{Q_1^2}{3}, \\
b_4 &=& \frac{Q_4}{Q_1} - Q_3 - \frac{Q_2^2}{2Q_1} + Q_2 Q_1 - \frac{Q_1^3}{4},
\eea
and so forth. The above equations are entirely based on thermodynamics and valid for arbitrary 
interaction and spatial dimension.

The task of calculating $b_n$ has typically been equated
with solving the $n$-body problem, constructing the $Q_n$, and inserting those in the above equations.
It is therefore not surprising that second-order VEs are easily carried out, as all that is needed for $b_2$ is 
the solution to the two-body problem. In fact, formulas exist for $b_2$ for many cases, some of which we quote below, 
based on the celebrated Beth-Uhlenbeck result~\cite{BU}.
Obtaining $b_3$ and beyond, however, typically requires numerical methods (see e.g.~\cite{DrummondVirial2D,virial2D2,Doerte}).
Although the $b_n$ are a proxy for other quantities, their calculation has become an attractive challenge per se, 
especially in cases such as the unitary limit~\cite{ZwergerBook} (the universal limit of zero interaction range and infinite scattering length), 
where the $b_n$ represent universal constants of quantum many-body physics. For that reason, the calculation of the $b_n$ has been 
vigorously pursued by several groups~\cite{LeeSchaeferPRC1,LiuHuDrummond,Leyronas,DBK,Rakshit,Ngampruetikorn,Doerte}.

In this work we focus on the virial coefficients of the generic lattice Hamiltonian of two-species nonrelativistic fermions with zero-range interactions, i.e.
\beq
\label{Eq:H}
\hat H = \sum_{\bf p} \frac{{\bf p}^2}{2m}\hat n_{{\bf p}} - g \sum_{\bf x} \hat n_\uparrow ({\bf x})  \hat n_\downarrow({\bf x}),
\eeq
where the total density operator in momentum space is $\hat n_{{\bf p}} = \hat n_{\uparrow, {\bf p}} + \hat n_{\downarrow,{\bf p}}$,
and $\hat n_s({\bf x})$ is the density for spin $s$ at position $x$. We will use units such that $\hbar = k_B = m = 1$.

For the above Hamiltonian, we obtain the first six virial coefficients of the one-dimensional (1D) case, i.e. the Gaudin-Yang 
model~\cite{GY}, and the first four virial coefficients of the two-dimensional (2D) case.
While the former is a classic problem that has been extensively studied (see e.g.~\cite{BA} for a recent review of 1D Fermi gases),
to our knowledge its virial coefficients beyond $b_2$ have not been calculated.
The 2D case, in contrast, has been under intense scrutiny in recent years, as it has been realized experimentally with ultracold atoms by several 
groups~\cite{Experiments2D2010,Experiments2D2011,Experiments2D2012,ContactExperiment2D2012,RanderiaPairingFlatLand,Experiments2D2014,Vale2Dcriteria}. 
Moreover, its thermal properties have been explored theoretically as well by various authors (see Ref.~\cite{Theory2D} for a review) 
and its virial coefficients $b_2$ and $b_3$ have been known for a few years.

To determine $b_n$, we developed two stochastic methods which bypass the direct solution of the $n$-body problem.
One of our objectives is to show that it is possible to design methods that allow to calculate high-order 
virial coefficients without solving the $n$-body problem, at the price of reduced precision.
The first method is based on the idea of Fourier particle-number projection of nuclear physics~\cite{NuclearParticleProjection},
as applied to the auxiliary field path-integral representation of $\mathcal Z$. That approach naturally yields a complex measure, 
and for that reason 
we implement the complex Langevin algorithm to sample the field~\cite{CL1}. The resulting method is able to compute high-order virial coefficients
at weak couplings and can also estimate the radius of convergence $\alpha_0$ of the VE as a function of the coupling strength. 
The second method consists in the stochastic evaluation of the {\it change} in the virial coefficients 
due to interaction effects, $\Delta b_n$. This second method uses the definition of the $b_n$ in their path-integral form derived 
from $\mathcal Z$, but it does not use $\mathcal Z$ directly. Thus, it is able to evaluate $b_n$ at stronger couplings than the 
projection method, but gives no information about the radius of convergence. Besides those two stochastic methods,
we implement a semiclassical lattice approximation (SCLA) at leading order (LO).
In all cases we use the known results for $\Delta b_2$ as the renormalization condition that connects the bare lattice coupling
to the physical coupling. 

The generalization of our approaches to higher dimensions is straightforward. In fact, 
the generic system studied here (a nonrelativistic gas with zero-range interactions) has been under 
intense investigation both theoretically and experimentally in the last decade in 1D, 2D, and 3D, 
and analytic results exist for $b_2$ in all dimensions based on the Beth-Uhlenbeck formula mentioned 
above~\cite{BU, EoS1D, virial2D, Daza2D, LeeSchaeferPRC1}.

{\it Formalism: Stochastic methods}.-
Using Eq.~(\ref{Eq:logZexpansion}), the $b_n$
can be obtained by Fourier projection. Following that route, we define the function
\beq
b_n(\alpha) \equiv 
\frac{1}{Q_1}\int_0^{2\pi}\frac{d\phi}{2\pi} e^{i \phi n}\ln \mathcal Z [z \to \alpha e^{-i\phi}] = b_n \alpha^n.
\label{Eq:bn}
\eeq
%
To proceed, we write $\mathcal Z$ as a path integral over a Hubbard-Stratonovich (HS) field $\sigma$
(see e.g.~\cite{MCReview,HSLee2}),
$
\label{Eq:PathIntegralZ}
\mathcal Z = \int \mathcal D \sigma\, {\det}^2 M[\sigma,z],
$
where we focus on unpolarized systems, thus the power of 2. 
The matrix $M[\sigma,z]$ encodes the dynamics and parameters of the system of interest; 
in particular, the $z$ dependence appears as $M[\sigma,z] = \openone + z U[\sigma]$,
where $U[\sigma]$ contains the kinetic energy and interaction information (see~\cite{MCReview} 
for details on the specific form of $M[\sigma,z]$ and $U[\sigma]$). Setting $z \to \alpha e^{-i\phi}$ and differentiating 
both sides with respect to $\alpha$ yields
\beq
\label{Eq:EVbn}
b_n 
=
\frac{1}{n \alpha^{n-1}}\frac{1}{Q_1}\!\int_0^{2\pi}\frac{d\phi}{2\pi} e^{i \phi n}
\langle \tr \left[2 M^{-1} \partial M/\partial\alpha\right]\rangle_{\phi,\alpha},
\eeq
where $P[\sigma,z]\equiv {\det}^2 M[\sigma,z]  / \mathcal Z [z]$,
and we have used angle brackets as a shorthand notation for the expectation value with $P[\sigma,\alpha e^{i\phi}]$
as a weight.
In practice, we use a discrete Fourier transform such that
\beq
\label{Eq:EVbnDiscrete}
\frac{\partial b_n(\alpha)}{\partial \alpha} = 
\frac{1}{ Q_1}\frac{1}{N_k} \sum_{k=0}^{N_k-1} e^{i \phi_k n}
\langle \tr \left[2 M^{-1} \partial M/\partial\alpha\right]\rangle_{\phi_k,\alpha}.
\eeq
where $\phi_k = 2\pi k /N_k$, $k = 0,\dots,N_k-1$, and $N_k$ is the number of discretization points.
This is the fundamental equation of the proposed approach. Calculating the expectation values inside the sum
in Eq.~(\ref{Eq:EVbnDiscrete}) for $N_k$ values
of $\phi_k$, and carrying out the Fourier sum for different values of $n$, one obtains the desired $b_n$.
In such a calculation, the results for $b_n$ must be independent of $\alpha$, such that that variable can be used
as a measure of the reliability of the method. In practice we plot
\beq
\label{Eq:bnalpha}
b_n = \frac{1}{n \alpha^{n-1}}\frac{\partial b_n(\alpha)}{\partial \alpha},
\eeq
as a function of $\alpha$ and fit a constant. 
The $\alpha^n$ dependence of the $n$-th order term is the main limiting factor in extracting 
high-order virial coefficients. To overcome that limitation, it is desirable to make $\alpha$ as large as possible but 
less than unity to remain in the virial region. Thus, deviations in Eq.~(\ref{Eq:bnalpha}) from constant behavior as $\alpha$ 
is {\it decreased} are indicative of uncertainties due to statistical noise or insufficient Fourier points.
On the other hand, non-constant behavior as $\alpha$ is {\it increased} indicates the appearance
of roots of $\mathcal Z$ in the complex-$z$ plane, which yield branch-cut singularities in $\ln \mathcal Z$ 
and point to the radius of convergence of the VE (see Supplemental Materials).

Evaluating the expectation values in Eq.~(\ref{Eq:EVbnDiscrete})
involves calculations that suffer from a phase problem, as $P[\sigma,\alpha e^{-i \phi}]$
will generally be a complex weight. To address that issue, we turn to complex stochastic quantization
via the complex Langevin (CL) method, which has recently been applied to the characterization 
of other aspects of non-relativistic fermions~\cite{PRDLoheacDrut, PRDLukas1, PRDPolarized, PolarizedUFG}. We employ the 
CL method in the same way described in Ref.~\cite{PRDLoheacDrut} (where it was 
applied to address repulsive interactions), setting the fugacity to $z \to \alpha e^{-i \phi_k}$. The quantity in the 
expectation value appearing in Eq.~(\ref{Eq:EVbnDiscrete}), namely $\tr \left[M^{-1} \partial M/\partial\alpha\right]$, 
corresponds to the density of the system. Thus, the proposed approach effectively consists in the Fourier projection
of the virial coefficients from the density equation of state, which is reminiscent of other approaches such as
those of Refs.~\cite{DBK, Leyronas, Ngampruetikorn, Daza2D}.

Our second method calculates the interaction effects on $b_n$ using 
their definition in terms of path integrals, derived analytically from the path integral form of $\mathcal Z$.
In that formalism, the change in $b_n$ due to interactions is
\bea
\label{Eq:Db2Db3}
\Delta b_2 &=& \frac{\Delta Q_{1,1}}{Q_1}, \ \ \ \ \ \Delta b_3 = \frac{2 \Delta Q_{2,1}}{Q_1} - Q_1 \Delta b_2, \nonumber \\ 
\Delta b_4 &=& \frac{ 2 \Delta Q_{3,1} + \Delta Q_{2,2}}{Q_1}  - \frac{Q_1^{2}}{2} \Delta b_2 - \frac{Q_1}{2}  (\Delta b^2_2 + 2 \Delta b_3), \nonumber
\eea
where $Q_{m,n}$ is the partition function for $m$ particles of one species and $n$ of the other,
and $\Delta Q_{m,0}=0$ because we only have contact interactions.
The VE of the fermion determinant yields
\bea
\label{Eq:Q11Q21}
Q_{1,1} &=& \!\!\int \mathcal D \sigma \tr^2 U[\sigma],\\
2 Q_{2,1} &=& \!\!\int \mathcal D \sigma \tr^3 U[\sigma] \left(1 \!-\! \frac{\tr U^2[\sigma]}{\tr^2 U[\sigma]}\right), \nonumber \\
2 Q_{3,1} &=& \!\frac{1}{3}\!\int \mathcal{D} \sigma \tr^4 U [\sigma] \left(1 \!-\! \frac{3\tr U^2[\sigma]}{\tr^2 U[\sigma]} \!+\! \frac{2 \tr U^3 [\sigma]}{\tr^3 U[\sigma]} \right), \nonumber \\
Q_{2,2} &=& \!\frac{1}{4}\!\int \mathcal{D} \sigma \tr^4 U[\sigma] \left ( 1-\frac{\tr U^2[\sigma]}{\tr^2 U[\sigma]} \right )^2,\nonumber
\eea
and so on at higher orders. Inserting these expressions in Eq.~(\ref{Eq:Db2Db3}) (and their noninteracting versions) 
yields stochastic formulas for $\Delta b_n$. To evaluate those, we use the usual two-species action $S[\sigma,z] = -2\ln \det M[\sigma,z]$ 
to sample $\sigma$, and extrapolate the results to the $z = 0$ limit.
This method is similar in spirit to that of Ref.~\cite{Doerte}, but employs a field integral
representation instead of an integral over particle paths.

{\it Formalism: Semiclassical lattice approximation}.-
Using the formulas of Eq.~(\ref{Eq:Q11Q21}), it is possible to implement what we call the semiclassical
lattice approximation, in which we neglect the commutator of the kinetic energy matrix $T$ and the potential energy
matrix $V$ at leading order. Thus, the matrix $U[\sigma]$ becomes simply $U[\sigma] = e^{-\beta T} \mathcal V[\sigma]$,
where $\mathcal V[\sigma]$ encodes the specific form of the HS transformation. Such an approximation amounts to 
a coarse discretization of the imaginary-time direction, which nevertheless becomes exact in two different limits: $V \to 0$ and $T \to 0$. 
In between those limits, higher orders in the SCLA can be reached by using finer temporal meshes; we leave 
calculations beyond LO to future work. At LO, the path integrals can be carried out analytically:
\bea
\label{Eq:Db3Db4SCLA}
\Delta b_3 &=& -2^{1-d/2}\Delta b_2, \\
\Delta b_4 &=&  2(3^{-d/2} + 2^{-d-1}) \Delta b_2 \nonumber \\ 
&& + 2^{1-d/2} \left(2^{-d-1} -1\right)(\Delta b_2)^2,
\eea
where we present our results in terms of $\Delta b_2$ because we will use the exact $\Delta b_2$ as a renormalization condition.

{\it Results: Virial coefficients in 1D}.- 
\begin{figure}[t]
  \begin{center}
   \includegraphics[scale=0.7]{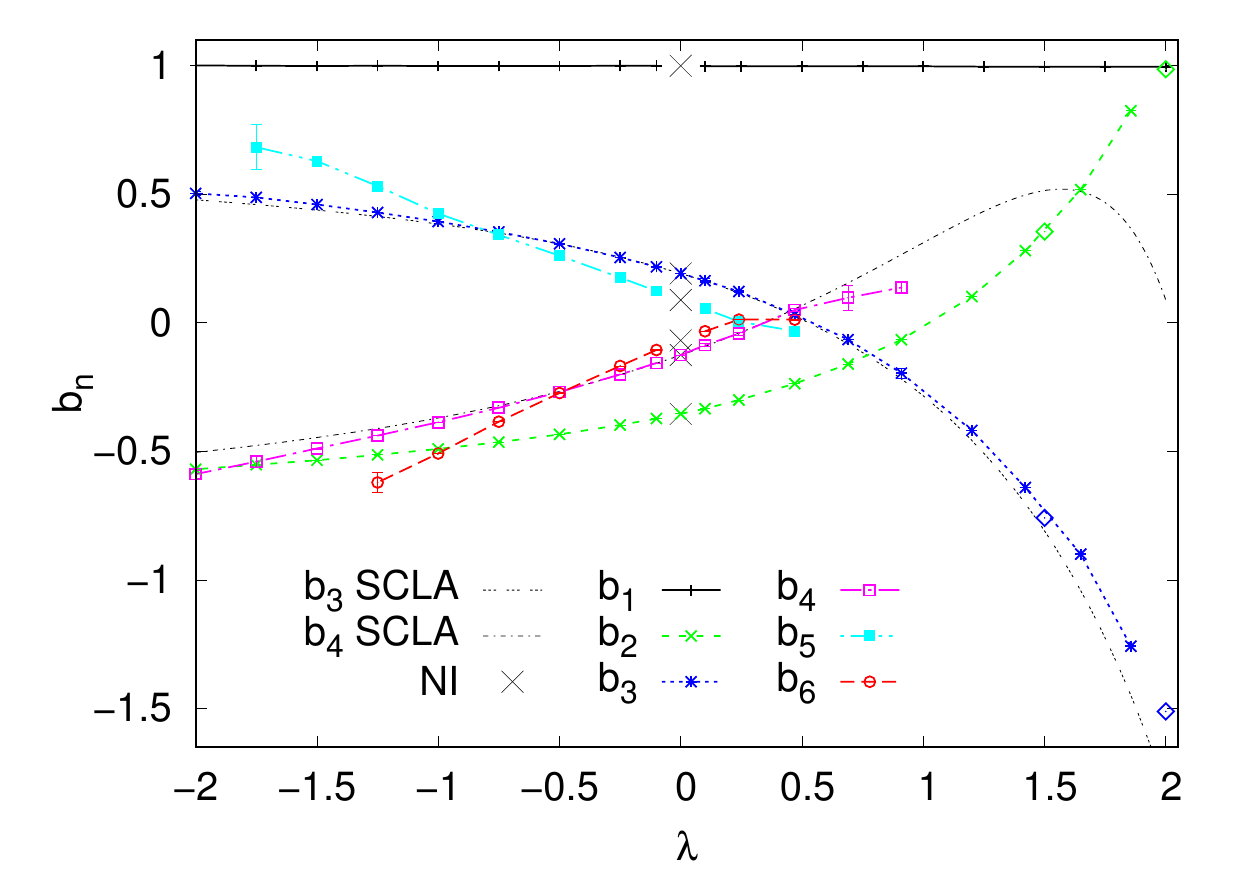}
  \end{center}
  \caption{Virial coefficients $b_n$ for $n=1-6$ for the 1D Fermi gas, as a function of the dimensionless coupling $\lambda$,
  as obtained with our projection method.
  Crosses on the $y$ axis denote the non-interacting values $b_n = (-1)^{n+1} n^{-3/2}$. 
  The leading order of the semiclassical lattice approximation
  (LO-SCLA) is shown with a dashed-dotted line for $\Delta b_3$ and with a dashed line for $\Delta b_4$.
   Green and blue diamonds show the results obtained with our second stochastic method, for comparison.
   }
  \label{Fig:LambdaDependence1D}
\end{figure}
To analyze the 1D case, our calculations used a lattice of spatial size $N_x = 30$ and temporal size
$N_\tau = 120 - 200$. We otherwise used the same lattice parameters as those of Ref.~\cite{PRDLoheacDrut}.
The number of Fourier points was set to $N_k = 30$ for the main results, with explorations covering
$N_k = 20-100$ showing no significant variation. By definition, $b_1=1$ and, for the 1D contact interaction studied here 
(see Ref.~\cite{EoS1D}),
\beq
\label{Eq:b2_1D}
b_2^\text{(1D)} = -\frac{1}{\sqrt{2}} + \frac{e^{\lambda^2/4}}{2\sqrt{2}}[1 + \text{erf}(\lambda/2)],
\eeq
where erf is the error function and $\lambda$ is the dimensionless coupling.
The noninteracting limit is $b_2^\text{(1D)} \to -\frac{1}{2\sqrt{2}}$.
We will use the analytic form of Eq.~(\ref{Eq:b2_1D}) as a renormalization condition, i.e. to {\it define} the
coupling $\lambda$ from our lattice determination of $b_2$. As a consequence, our plots of $b_2$
below will be exact by definition.
Our first result appears in Fig.~\ref{Fig:LambdaDependence1D}, where we map out the $\lambda$ dependence of the first six $b_n$. 
The smoothness of the results gives confidence that the method works as expected.
Perhaps the most prominent feature in Fig.~\ref{Fig:LambdaDependence1D} is the monotonicity 
of the stochastic data for each $b_n$: besides the constant $b_1 = 1$, the even $n$ coefficients
increase as a function of $\lambda$, whereas the odd ones decrease.
More specifically, toward the repulsive side ($\lambda < 0$), the $b_n$ grow in magnitude and maintain their sign: the even ones
which start out negative at $\lambda = 0$ become more negative and the odd ones which start positive grow as well. 
Toward the attractive side, the monotonic behavior implies that in a wide region $0 < \lambda < 1$ many of the coefficients
cross the $b_n= 0$ line, which suggests the VE may be useful up to $z \simeq 0.5$ (see however
our results below for the radius of convergence). Beyond that point,
the coefficients grow in magnitude and eventually change sign relative to their noninteracting values.
Using the second stochastic method (applied below in 2D), we checked the above results of Fig.~\ref{Fig:LambdaDependence1D} for $b_2$ and $b_3$.

\begin{figure}[t]
  \begin{center}
   \includegraphics[scale=0.72]{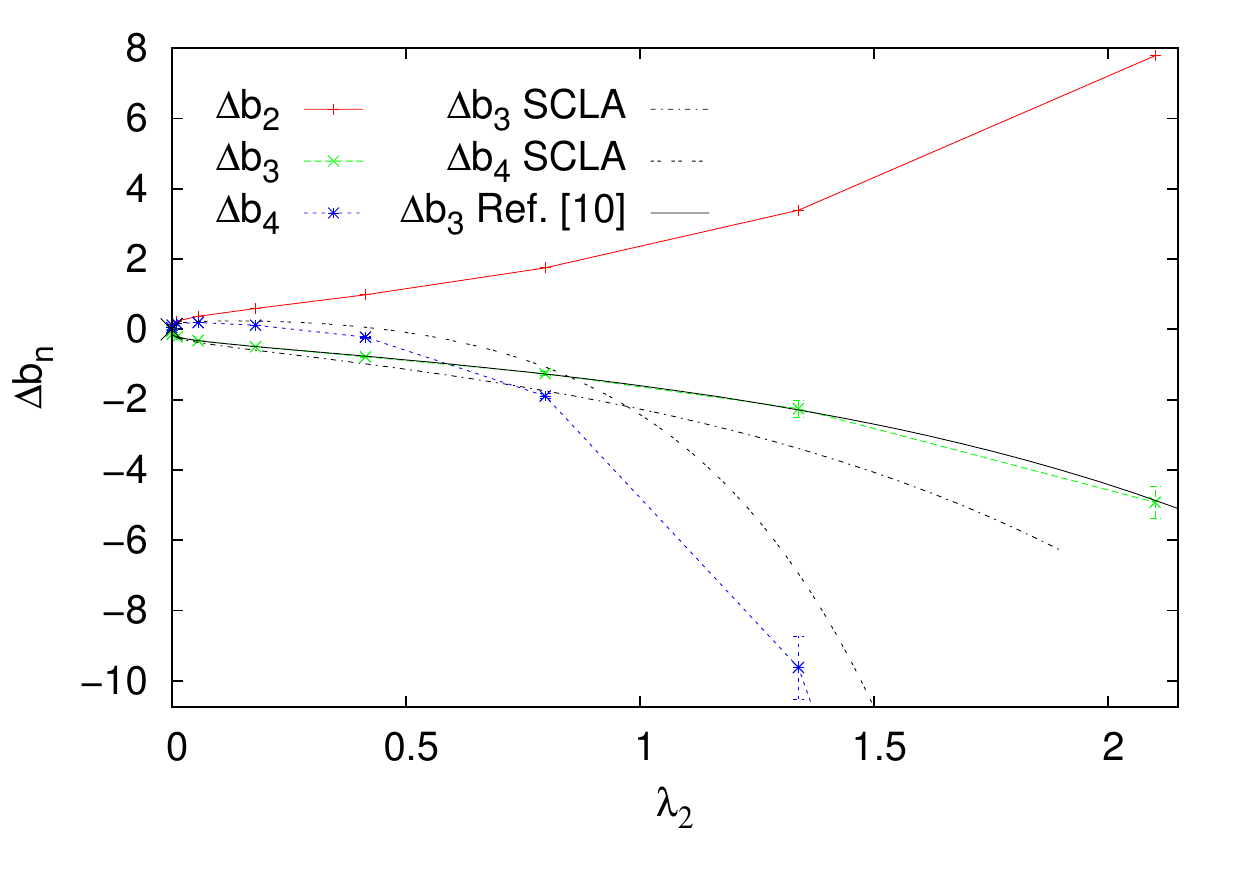}
  \end{center}
  \caption{Interaction change of the virial coefficients $\Delta b_n$ for $n=2-4$ for the 2D Fermi gas, 
  as a function of the dimensionless coupling $\lambda_2$. The solid red line connects the data for $\Delta b_2$,
  the green shows $\Delta b_3$, and the blue shows $\Delta b_4$. The leading order of the semiclassical lattice approximation
  (LO-SCLA) is shown with a dashed-dotted line for $\Delta b_3$ and with a dashed line for $\Delta b_4$.
  The solid black line shows the result for $\Delta b_3$ of Ref.~\cite{virial2D2}.
  Note that the data for $\Delta b_2$ reproduces the exact result of Eq.~(\ref{Eq:b22D}) by virtue of the renormalization condition (see text).
  }
  \label{Fig:LambdaDependence2D}
\end{figure}

{\it Results: Virial coefficients in 2D}.- 
Besides the 1D case above, we applied the second method to the 2D analogue,
which was studied up to second order in the VE in Refs.~\cite{virial2D,Ordo,Daza2D} and up to third order in 
Refs.~\cite{DrummondVirial2D, virial2D2}.
The Hamiltonian is essentially identical to that of Eq.~(\ref{Eq:H}), generalized to 2D.
In that case, the coupling $g$ becomes simply a bare parameter and the physical coupling 
is given by $\lambda_2 = \sqrt{\beta \varepsilon^{}_B}$, where $\varepsilon^{}_B$ is the binding 
energy of the two-body system. The second-order virial coefficient in 2D is known~\cite{virial2D, Daza2D, EoS2D} 
and given by
\beq
\label{Eq:b22D}
b_2^\text{(2D)} = -\frac{1}{4} + e^{\lambda_2^2} - \int_0^{\infty} \frac{dy}{y} \frac{2 e^{-\lambda_2^2 y^2}}{\pi^2 + 4 \ln^2 y}.
\eeq
The noninteracting limit yields $b_2^\text{(2D)} \to -\frac{1}{4}$.
As in our 1D calculations, we used Eq.~(\ref{Eq:b22D}) to define $\lambda_2$ by calculating
$b_2$ on the lattice.
In Fig.~\ref{Fig:LambdaDependence2D} we show our results for $b_2$, $b_3$, and $b_4$.
By definition, $b_2$ is reproduced exactly, and the output of the calculation is $b_3$ and $b_4$.

{\it Results: Semiclassical lattice approximation}.-
The predictions of the LO-SCLA are compared with those of our stochastic methods in Figs.~\ref{Fig:LambdaDependence1D}
and~\ref{Fig:LambdaDependence2D}. The LO-SCLA predicts in 1D: $\Delta b_3 = -\sqrt{2} \Delta b_2$ 
and $\Delta b_4 = (4\sqrt{3} + 3)/6\,\Delta b_2 - 3\sqrt{2}/4\,(\Delta b_2)^2$; and in 2D: $\Delta b_3 = -\Delta b_2$ and 
$\Delta b_4 =  11/12\,\Delta b_2 -  7/8\,(\Delta b_2)^2$. 
As is clear in Figs.~\ref{Fig:LambdaDependence1D} and~\ref{Fig:LambdaDependence2D},
there are differences between those predictions and the stochastic results. However, it is remarkable that 
at LO the SCLA predicts not only the correct sign of $\Delta b_3$ but also a deviation smaller than $10\%$ in 1D and close 
to $20\%$ in 2D, at least for the regime of couplings that studied here. Such results encourage higher orders studies of the SCLA,
which will be carried out elsewhere.

While we focus here on 1D and 2D, it is also interesting to test the predictions of the LO-SCLA 
for the 3D Fermi gas at unitarity.
There, known results (see e.g.~\cite{LiuHuDrummond, Leyronas, DBK, Rakshit, Ngampruetikorn}) give 
$\Delta b_2 = 1/\sqrt{2}$ and
$\Delta b_3 = -0.35505\dots$, such that $\Delta b_3/\Delta b_2 \simeq -0.50\dots$, while the LO-SCLA yields
$\Delta b_3/\Delta b_2 = -1/\sqrt{2} \simeq -0.707$, thus matching the correct sign of $\Delta b_3$ but overshooting 
its magnitude by about $40\%$.
Similarly, the most accurate result at unitarity~\cite{Doerte} is $b_4 = 0.078(18)$, which yields $\Delta b_4 = 0.109(18)$, 
while the LO-SCLA yields $\Delta b_4 =  0.029\dots$,
which matches the sign of the expected result but undershoots its magnitude by roughly a factor of 3. 
Nevertheless, these results are encouraging when considering that they come from a mere leading-order approximation.

{\it Summary and Conclusions}.-
We have calculated the first few virial coefficients $b_n$ of two systems: fermions in 1D and 2D, 
both with a contact interaction. In 1D, we evaluated the first six $b_n$ as a function 
of the coupling strength $\lambda$ in both attractive and repulsive regimes. 
In the 2D case, we calculated $\Delta b_3$, and $\Delta b_4$ for attractive interactions. 
To carry out our calculations, we implemented two different stochastic lattice methods. 
The first method relied on projecting the $b_n$ out of the path integral form of the density equation of
state. The second approach used a path-integral representation of the virial coefficients,
as derived from the path integral form of $\mathcal Z$. The 
latter method enables calculations in a way that  
requires neither matrix inversion nor determinants, but which is sensitive to statistical noise 
as $n$ is increased, due to the various volume-scaling cancelations required to resolve each $b_n$
from the canonical partition functions. However, that noise can at least partially be addressed by obtaining 
more samples, a task that can be carried out in a perfectly scalable fashion. The stochastic approaches proposed 
here are not as precise as exact diagonalization, but provide a systematic way to high-order coefficients without
solving the $n$-body problem. Finally, we used a semiclassical approximation which at leading order compares
remarkably well with our stochastic results for the coupling strengths studied.

\acknowledgments
{
We thank A. C. Loheac for discussions and for the perturbation theory data of Ref.~\cite{PRDLoheacDrut},
W. J. Porter for discussions in the early stages of the project, and J. Levinsen and C. Ord\'o\~nez for comments 
on the manuscript. We also thank the authors
of Ref.~\cite{virial2D2} for kindly providing us with their data for $\Delta b_3$ in 2D.
This material is based upon work supported by the National Science Foundation under Grant No.
PHY{1452635} (Computational Physics Program). J.E.D would like to acknowledge the hospitality
of the Institute for Nuclear Theory at the University of Washington, where part of this work was
carried out.
}



\section{Supplemental Materials for \\ ``Virial coefficients of 1D and 2D Fermi gases by stochastic methods \\ and a semiclassical lattice approximation''}

\subsection{Radius of convergence via projection method}
\begin{figure}[b]
  \begin{center}
   \includegraphics[scale=0.7]{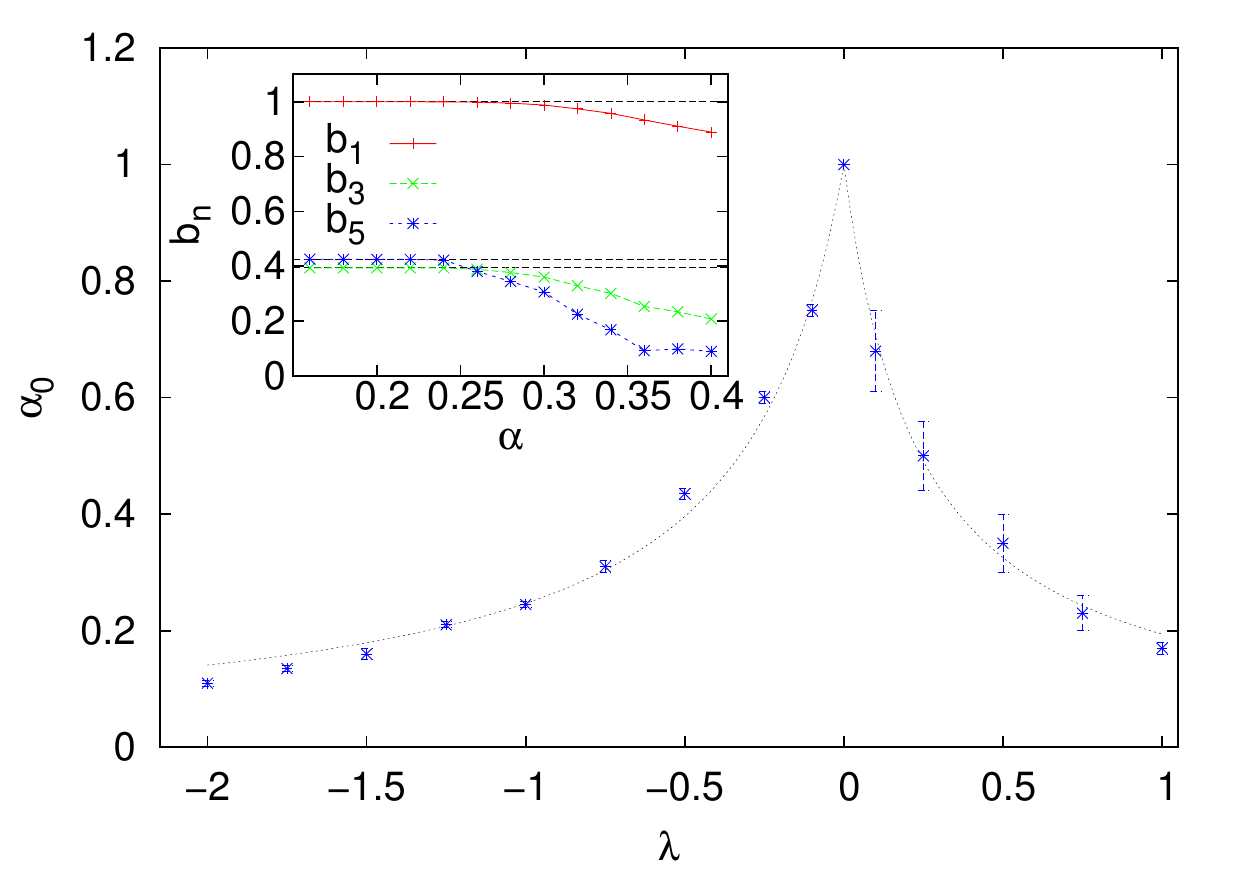}
  \end{center}
  \caption{Estimate of the radius of convergence $\alpha_0$ of the virial expansion as a function of the coupling $\lambda$.
 Inset: $b_n$ for $n=1,3,5$ for $\lambda=-1$. Constant behavior as a function of $\alpha$ is expected when the coefficient of the $n$-th
 power of $z$ is extracted successfully. Deviation from such a constant as $\alpha$ is increased shows the appearance of 
 roots of $\mathcal Z$ in the complex-$z$ plane, which yields the estimate $\alpha_0$ for the radius of convergence shown in the main plot.}
  \label{Fig:XiDependence}
\end{figure}

In the inset of Fig.~\ref{Fig:XiDependence} we show $b_n$ as a function of $\alpha$ (see main text). 
As anticipated, for each virial order $n$ there is a region around $\alpha=0$ for 
which $b_n$ does not vary, which allows us to extract the value of $b_n$ itself.
Beyond a $\lambda$-dependent value of $\alpha$, however, the calculation runs into the roots of $\ln \mathcal Z$
in the complex plane and the constant behavior is lost. We stress that this is not due to systematic or 
statistical effects, but rather a feature of the calculation that represents the radius of convergence $\alpha_0$ of the virial expansion.
The main plot of Fig.~\ref{Fig:XiDependence} shows our results for $\alpha_0$ as a function of $\lambda$, obtained by locating
the point where the constant behavior as a function of $\alpha$ is lost. Our results are consistent with the expected value 
$\alpha_0=1$ for the noninteracting case, which is easily derived by noting that the noninteracting partition function has a root at $z=-1$.
The dashed line in the main plot of Fig.~\ref{Fig:XiDependence} shows a fit $\alpha_0(\lambda) = 1/(1+ C|\lambda|)$, where
$C \simeq 3.05(5)$ on repulsive side ($\lambda < 0$) and $C \simeq 4.15(5)$ on attractive side ($\lambda > 0$).
While the fit is merely descriptive, it does point to
a nontrivial feature, namely the non-analyticity of $\alpha_0$ around the maximum at $\lambda=0$: the data appears to display a cusp.

\subsection{Semiclassical approximation}
From the equations in main text it is easy to see that
\beq
\Delta b_2 = \frac{\Delta Q_{1,1}}{Q_1} = 
\frac{1}{Q_1} \int \mathcal D \sigma \left( \tr^2 U[\sigma] - \tr^2 U_0 \right),
\eeq
where $U_0 = e^{-\beta T}$ is the noninteracting transfer matrix ($T$ being the kinetic energy matrix),
and $U[\sigma]  = e^{-\beta T} \mathcal V [\sigma]$ ($\mathcal V$ being the chosen Hubbard-Stratonovich representation
of the interaction). Carrying out the path integrals, it is straightforward to find
\beq
\Delta b_2 = (e^{\beta g} - 1) \frac{V}{Q_1} \left( \frac{\tr U_0}{V}\right)^2
\eeq
where $Q_1/V \to 2/\lambda_T^d$ in the continuum limit in $d$ spatial dimensions and all lengths
are in units of the lattice spacing $\ell = 1$. Moreover, $\tr U_0= Q_1/2$, such that, in the continuum limit,
\beq
\Delta b_2 = \frac{1}{\lambda_T^d} \frac{e^{\beta g} - 1}{2}.
\eeq
The calculation of $\Delta b_3$ is only slightly more tedious and yields
\beq
\Delta b_3 = \frac{2\Delta Q_{1,1}}{Q_1} - Q_1 \Delta b_2= - \frac{1}{\lambda_T^d} \frac{e^{\beta g} - 1}{2^{d/2}}.
\eeq
We thus obtain the result advertised in the main text, namely
\bea
\Delta b_3 &=& -2^{1-d/2}\Delta b_2.
\eea
The calculation of $\Delta b_4$ follows the same steps but yields a contribution that is quadratic in $\Delta b_2$:
\bea
\Delta b_4 &=&  2(3^{-d/2} + 2^{-d-1}) \Delta b_2 \\ 
&& + 2^{1-d/2} \left(2^{-d-1} -1\right)(\Delta b_2)^2.
\eea

\subsection{Systematic effects}

Because we chose a lattice regularization to carry out our calculations, 
there are a few systematic effects that need to be taken into account. 
First of all, we have put the system on a lattice and must describe how to take the continuum limit. 
That amounts to enlarging the window
$\ell \ll \lambda_T \ll L$, where $\ell = 1$, $L = N_x \ell$, 
and $\lambda_T = \sqrt{2\pi \beta}$ is the thermal wavelength.

\begin{figure}[h]
  \begin{center}
   \includegraphics[scale=0.7]{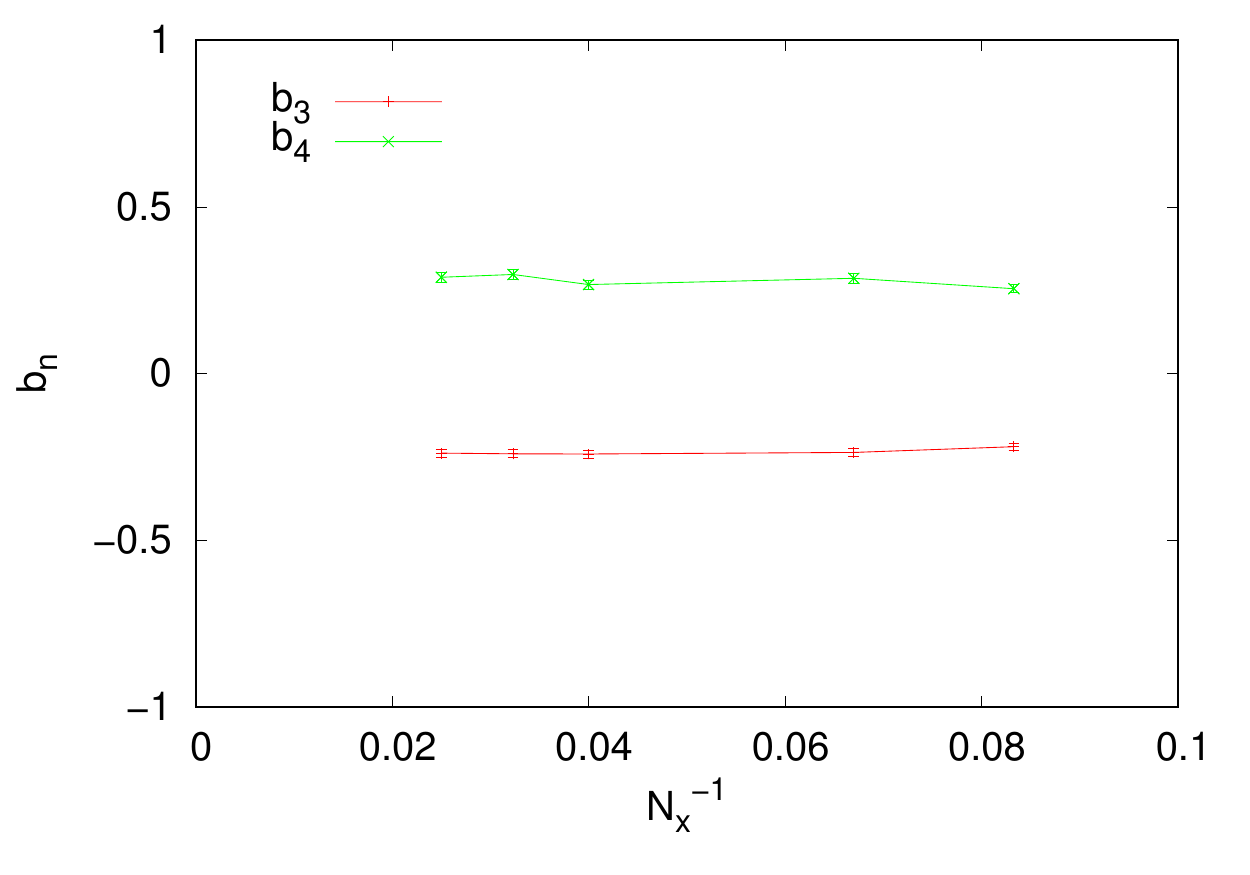}
      \includegraphics[scale=0.7]{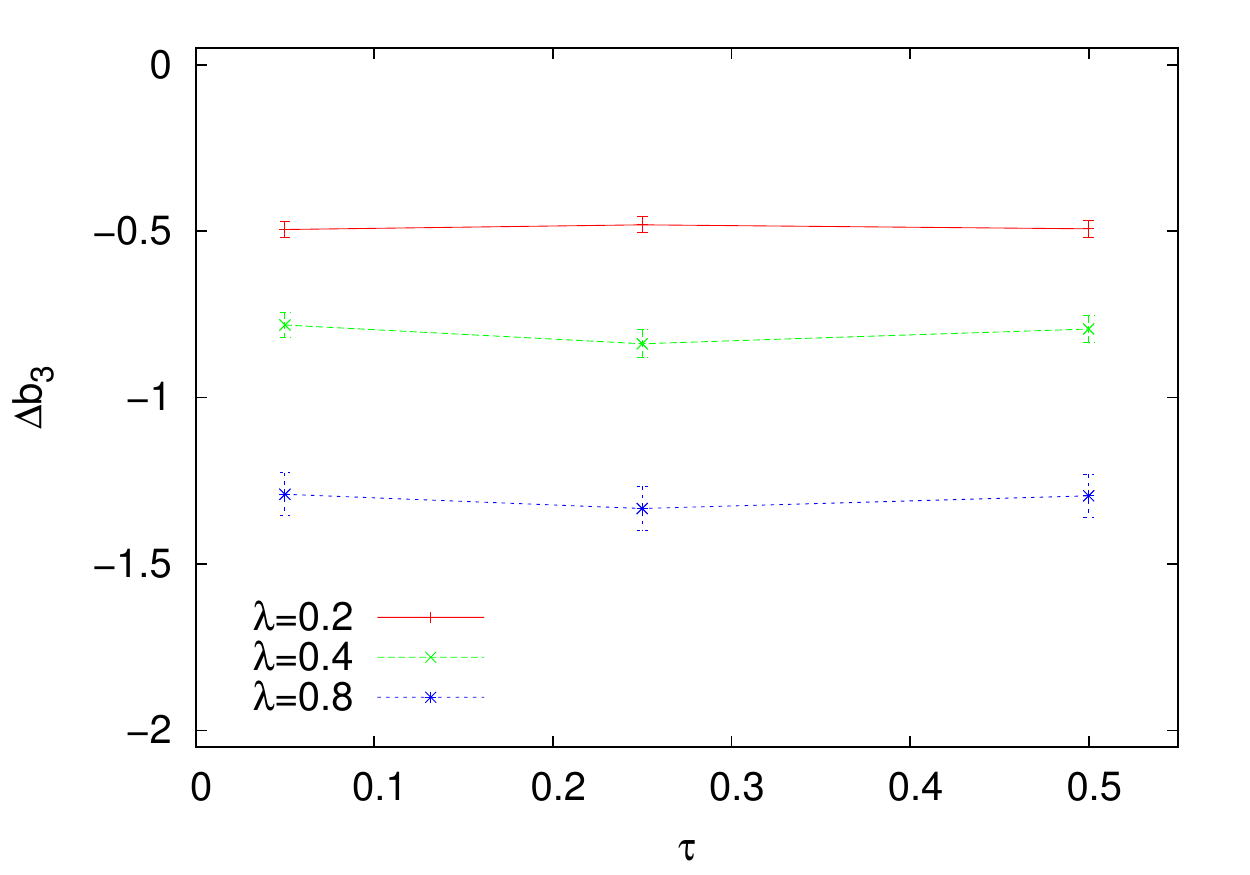}
  \end{center}
  \caption{Top: Illustration of the size of the finite-$N_x$ effects on $b_3$ and $b_4$ in 1D at $\lambda = 1$. The errorbars show statistical effects. Bottom: Illustration of the size of the finite-$\tau$ effects on $\Delta b_3$ in 2D for varying $\lambda$.}
  \label{Fig:NxTauDependence}
\end{figure}

Our main results correspond to $N_x = 30$ and $\lambda_T \simeq 7$, such that the above
window is well satisfied. As an illustration of the size of the finite-$N_x$ effects, we show results for varying 
$N_x$ in Fig.~\ref{Fig:NxTauDependence} (top). The variation is appreciable but small on the scale of the corresponding
plot in the main text.
\newline

\indent The second systematic effect to account for is the number of Fourier points $N_k$ used
for the projection. Relying on Nyquist's theorem,
taking $N_k$ at least twice as large as the highest desired virial coefficient $n_\text{max}$
should be sufficient. However, that lower bound turns out to be much too optimistic in practice. 
As a conservative choice, we set $N_k = 30$ and find that it enables projections up to $n=6$ 
with up to two decimal places. Note that the computation time scales linearly with $N_k$ and is 
perfectly parallelizable in that variable. 
\newline
\indent The third systematic effect is the dependence on the temporal lattice spacing $\tau$. We have tested 
$\tau = 0.05$, $0.25$, and $0.5$, as shown in Fig.~\ref{Fig:NxTauDependence} (bottom). Remarkably, the 
variation is small on the scale of the plot in the main figure (somewhat zoomed-in here).


\end{document}